\documentclass[prb,showpacs,twocolumn]{revtex4-1}%
\usepackage{amsfonts}
\usepackage{amsmath}
\usepackage{amssymb}
\usepackage[dvips]{graphicx}%
\usepackage{float}
\usepackage{epsfig}
\usepackage{subfigure}

\begin{document}

\title{\textbf{RSC Adv. 4, 29884(2014)}\\Spin-polarized surface state in Li-doped SnO$_{2}$(001)}%

\author{Naseem Ud Din and Gul Rahman}%
\email{gulrahman@qau.edu.pk}

\affiliation{Department of Physics, Quaid-i-Azam University Islamabad 45320 Pakistan}

\begin{abstract}

Using LDA+$U$, we investigate Li-doped rutile SnO$_2$(001) surface. The surface defect formation energy shows that it is easier for Li to be doped at surface Sn site than bulk Sn site in  SnO$_2$.  Li at surface and sub-surface Sn sites has a magnetic ground state, and the induced magnetic moments are not localized at Li site, but spread over Sn and O sites. The surface electronic structures show that Li at surface Sn site shows 100\% spin-polarization (half metallic), whereas Li at sub-surface Sn site does not have half metallic state due to Li-Sn hybridized orbitals. The spin-polarized surface has a ferromagnetic ground state, therefore, ferromagnetism is expected in Li-doped SnO$_2$(001) surface.

\end{abstract}
\maketitle

In the past, decade density functional theory (DFT) has proven to be a predictive  tool to discover new materials for certain applications, specially in the area of magnetism. With DFT, many new materials have been discovered and then synthesised.~\cite{1a, KCa, Nit, Mg,zn} 
DFT has also predicted spin polarized materials\cite{gr6,gr7,grprb}.
One of the new materials is oxide-based diluted magnetic semiconductor, which has potential applications in spintronics. The main quest in this area is to discover magnetic materials having transition temperature ($T_c$), which is the temperature at which a system changes from a paramagnetic(disorder phase) to a magnetic phase(order phase), well above room temperature and large magnetization and spin-polarization. With this hope, transition-metals (TMs) were doped into nonmagnetic (NM) semiconductor hosts,~\cite{1,2} but later on these TM doped systems were found to have inherent issues, i.e., clustering, antisite defects.~\cite{cluster}

SnO$_2$-based diluted system evoked particular attention  when S. B. Ogale \textit{et al.}~\cite{3} found a giant magnetic moment (GMM) in Co-doped SnO$_2$. Following this discovery, TM doped-SnO$_2$ has been extensively studied both experimentally and theoretically.~\cite{4,5,6,7,8,9,10} Later on in 2008, our theoretical  calculations showed that the Sn vacancies are responsible for magnetism in SnO$_2$.~\cite{11} This opened a new area of magnetism, where magnetism is made possible without doping of magnetic impurities, which are confirmed experimentally.~\cite{01,02,ZnOprl} 
To go beyond vacancy-induced magnetism, we also proposed possible magnetism induced by light elements, e.g., C, and Li.~\cite{apl, arxi} Recent theoretical calculations further show that magnetism can be induced with NM impurities.~\cite{KCa, Nit, Mg,12,13} A good example of NM impurity is carbon, which has been shown theoretically and experimentally that C-doped SnO$_2$ films can exhibit feromagnetic behaviour at room temperature,~\cite{apl,exp} where C does not induce magnetism in bulk SnO$_2$, when located at the oxygen site.~\cite{apl,exp}
Now, it is a firm belief that magnetism in NM hosts can be tuned either by vacancies or light elements. In oxides, the magnetic vacancies can be created either at cation site or anion site. Most of the theoretical work show that the cation vacancies are magnetic,~\cite{hf,hf2, tio2, ZnO} but there has remained an open question that how to stabilize magnetic vacancies due to their higher formation energies? Very recently, this issue is also addressed and we have demonstrated that doping of non magnetic specie (Li) can stabilize the intrinsic defects in SnO$_2$ appreciablly and also polarizes the host bands to induce magnetism in bulk SnO$_2$.~\cite{arxi} A very recent experimental report on nanoparticles of Li-doped SnO$_2$ also shows ferromagnetism and the XRD shows that Li is substituted at Sn site.~\cite{SnO2-Nano} 
In this article, we are mainly interested in the thermodynamic stability of Li-doped SnO$_2$ surface because  nanoparticle has a large surface to volume ratio and it is expected that the magnetism of nonparticles will be mainly governed by the surface properties. We are also looking for spin-polarized surface state, which is the key ingredient for
the special behavior of three-dimensional materials. \cite{hasan,xia}
Practical advantages of spin-polarized surfaces are in the field of spintronics,
where thin films are used as a spin filter material that shows
a high degree of spin polarization\cite{Santos}. Therefore, in thin films the electronic  and magnetic properties of the surface can play a significant role.
 
To study surface magnetism, we performed calculations in the framework of density functional theory (DFT), ~\cite{DFT} using linear combination of atomic orbital (LCAO) basis as implemented in the SIESTA code ~\cite{siesta}. A double-$\zeta$ polarized (DZP) basis set for all atoms was used, which included $s$, $p$ and $d$ orbitals in Sn and O (we polarized $p$ orbitals, which added additional 5 $d$ orbitals) and $s$ and $p$ orbitals
in Li (we polarized an $s$ orbital, which added 3 $p$ orbitals). These used basis sets are well tested in our previous work,\cite{11} where we found a good agreement with the FLAPW code. Such agreement insures the quality of the basis sets used in the present work. The local density approximation (LDA)~\cite{lda} is adopted for describing exchange-correlation interactions. We use standard norm-conserving pseudopotentials ~\cite{ps} in their fully nonlocal form ~\cite{pss}. Atomic positions and lattice parameters are optimized, using conjugate-gradient algorithm ~\cite{cg}, until the residual Hellmann-Feynman force on single atom converges to less then 0.05 eV/\AA. A cutoff energy of 400 Ry for the real-space grid was adopted. This energy cutoff defines the energy of the most energetic plane wave that could be represented on such grid, i.e. the larger the cutoff the smaller the separation between points in the grid ($E\sim G^2\sim 1/d^2$, where $\vec G$ is a reciprocal vector and $d$ is the separation between points). The sampling of $k$-space is performed with Monkhorst and Pack (MP) scheme with a regularly spaced mesh of  $5\times 5\times 1$ . Convergence with respect to $k$-point sampling was carefully checked. Our previous work shows that the magnetism of SnO$_{2}$ is not very sensitive to exchange correlation functionals.~\cite{11} Therefore, we only used LDA for Li-doped SnO$_{2}$ (001). Using the relaxed LDA atomic volume/coordinates, we also carried out LDA+$U$ calculations by considering the on-site Coulomb correction ($U$ = 6.0 eV, our previously optimized value\cite{arxi}) between the $p$-orbital electrons of O.~\cite{38,39} Note that the LDA$+U$ calculated band gap of SnO$_{2}$ in our case is $\sim 3.10 $ eV, which is comparable with the experimental and theoretical values of 3.20 eV\cite{iop,dean}. Generally, it may be difficult to see the direct effect of $U$ on magnetism from an experiment.  Indeed, many theoretical calculations have the same conclusion that LDA and LDA+$U$ almost give the same magnetism.  LDA /GGA or LDA/GGA+$U$ usually predicts a trend or possibility of magnetism.  For example, the LDA predicted possible magnetism of C-doped SnO$_{2}$ ~\cite{apl} is in good agreement with the experimental work ~\cite{exp}.  Therefore, we mainly used LDA+$U$ to predict the true impurity bands and  defect formation energies of Li-doped SnO$_{2}$ (001).

To address the thermodynamic stability of Li-doped SnO$_2$ (001) surface,  surfaces of SnO$_2$ with different number of layers are considered. Each layer is composed of SnO$_2$ surface unit. Free stoichiometric slabs with total compositions of Sn$_7$O$_{14}$ (seven layers), Sn$_9$O$_{18}$ (nine layers), and Sn$_{11}$O$_{22}$ (eleven layers), separated by a vacuum region of $\sim$ 10\;\AA, were constructed. Vacuum region is added so that the two surfaces do not interact with each other through the vacuum region. 
Our studied concentration of Li in SnO$_2$(001) is comparable with the recent experimental work, where  9.0\% of Li was doped into SnO$_{2}$  nanoparticles  and found to be FM.~\cite{SnO2-Nano} We further state increasing the Li concentrations in SnO$_{2}$ may have small effect on our theoretical results because at higher concentration of Li, the Li atom may occupy the interstitial site that may destroy magnetism in SnO$_{2}$.\cite{arxi,saif}

\begin{figure}[b]
\begin{center}
\includegraphics[totalheight=0.25\textheight ,
width=0.25\textwidth]{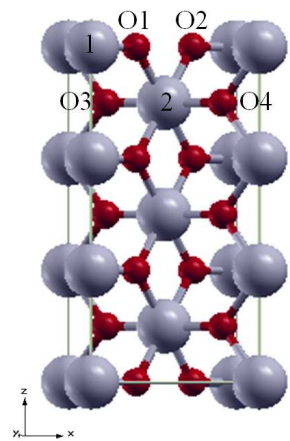}
\caption{Stoichiometric supercells of SnO$_2$ used in the calculations for the (001) surface. Big and
small balls represent Sn and O atoms, respectively. The surface O and Sn atoms are represented
by O1, O2 and Sn(1), respectively. The immediate sub-surface atoms are represented by O3, O4 and Sn(2).}
\label{surmodel}
\end{center}
\end{figure}

A representative (001) surface of SnO$_2$ is shown in Fig.~\ref{surmodel}, where both Sn and O are on the surface. In order to study the surface defect formation energy and effect of impurity on the surface magnetism and electronic structures, two types of systems were modelled: (a) Li doped at Sn(1) surface site and (b) Li doped at Sn(2) sub-surface site. 
Note that the unrelaxed atomic positions  of SnO$_{2}$  (001)are taken from our optimized structure of SnO$_{2}$.\cite{11,arxi} In the surface calculations, we relaxed all the atoms to find a minimum energy position. Such relaxation is essential to observe either surface reconstruction, which we did not observe, or saturate the dangling bonds.
Since the surface slabs have two dimensional periodic boundary conditions, the atomic positions shift along the surface ($xy$ ) plane should be small, which can be seen from our calculated results in Table~\ref{t1}. The significant shift is along the $z$ direction, which is assumed perpendicular to the surface plane. The surface oxygen atoms (O${1}$, O${2}$) relaxed in the upward direction due to surface strain induced by Li at Sn(1) site. On the other hand, the sub-surface oxygen atoms (O${3}$, O${4}$) relaxed inward, which is smaller than the surface O atoms. The Li atom was also relaxed in the outward direction. We must mention that such relaxation did not affect the formation energies and magnetism of Li-doped SnO$_2$.

\begin{table}
\caption{For a seven layered system, change in coordinates $\Delta{x}$, $\Delta{y}$, $\Delta{z}$, of surface, sub-surface atoms calculated as the difference of relaxed and unrelaxed coordinates in units of \AA. For $\Delta{z}$ values, the negative (positive) sign means upward (downward) surface relaxation.}
\begin{tabular}{llll}
\toprule
atom &$\Delta{x}$&$\Delta{y}$&$\Delta{z}$ \\
\colrule
Sn&0.00&0.00&0.01 \\
Li&0.00&0.00&0.30\\
O1&0.25&0.25&0.44 \\
O2&-0.25&-0.25&0.44\\
O3&0.01&-0.01&-0.11\\
O4&-0.01&0.01&-0.11\\
\botrule
 \end{tabular}
 \label{t1}
\end{table}

Once the optimized surface structure of SnO$_2$ (001) was determined, then we used LDA+$U$ to investigate the thermodynamic stability, magnetism, and electronic structures of SnO$_2$ systems.
The formation energies (E$_\mathrm{f}$) were calculated under three conditions; the equilibrium condition, O-rich condition and Sn-rich condition, as discussed in Ref.\cite{arxi}  
The surface defect formation energies by doping Li at the surface and sub-surface Sn sites were calculted using the following equation.
\begin{equation}
E_\mathrm{f}= E(\mathrm{Sn_{1-x}Li_{x}O_{2}})-E(\mathrm{SnO_{2}})+n \mu_\mathrm{Sn}-n \mu_\mathrm{Li} ,
\end{equation}
where $\mu_\mathrm{Li}$ is the chemical potential of Li calculated as total energy of bulk Li, $n$ is the number of atoms added or removed from host material, $E(\mathrm{Sn_{1-x}Li_{x}O_{2}})$ is total energy of Li-doped SnO$_2$ system and $E(\mathrm{SnO_{2}})$ is the total energy of pure SnO$_2$ system.
Table~\ref{113} lists the calculated surface defect formation energies. These calculated formation energies clearly suggest that Li doped at surface site Sn(1) has the lowest formation energy under equilibrium and O rich conditions. These values are much smaller than the bulk Li-doped SnO$_2$. The bulk values are 
-0.84 and 6.14 eV in E$_{\rm Sn}$ and E$_{\rm O}$ conditions, respectively\cite{arxi}. Interestingly, 
the formation energy for the case of Li doped at the sub-surface site Sn(2) is still smaller than the bulk case.
However, small changes are due to crystal environment as compared to surface Sn case. Now, doped Li is coordinated with six nearest O atoms. We repeated the same calculations for the nine and eleven layered systems (not shown here), and we got similar conclusion. These calculated results show that the number of layers (thickness of the films)  does not affect significantly the surface defect formation energy. From these thermodynamics, 
therefore, we conclude that it is easier for Li to be doped at surface Sn site than the bulk Sn site. We believe that Li can easily be doped at Sn site either in thin films or nanoparticles of SnO$_2$.  

\begin{table}[t]
\caption{Surface defect formation energies (in units of eV) of seven layered system, calculated under equilibrium (E$_\mathrm{eq}$), Sn-riched (E$_\mathrm{Sn}$) and O-riched (E$_\mathrm{O}$) conditions. In both surface and sub-surface Li substitutes Sn atom.  Values in parenthesis show formation energies calculated using LDA+$U$}
\begin{tabular}{lcll}
\toprule
System &E$_\mathrm{eq}$ &E$_\mathrm{Sn}$ &E$_\mathrm{O}$ \\
\colrule
Surface&-5.31(-7.42)&2.25(-0.44)&-5.31 (-7.42)\\
Sub-surface&-1.99(-2.71)&5.58 (4.27) &-1.99(-2.71)\\
\botrule
 \end{tabular}
 \label{113}
\end{table}

As bulk Li-doped SnO$_2$ shows magnetism, when Li is doped at Sn site\cite{arxi}, here we also investigate the possible surface magnetism of Li. It is encouraging that Li always shows magnetism when doped at surface Sn(1) or subsurface Sn(2) sites.
Table~\ref{LMM1} lists the local magnetic moments of Li, Sn, and O atoms when Li was doped at the surface Sn(1) and sub-surface Sn(2) sites
of seven layered system. When added atom Li goes to the surface site Sn(1), the magnetic moment induced on the each surface oxygen atom (O1,O2) is 1.05 $\mu_\mathrm{B}$, whereas the magnetic moment induced on the sub-surface O atom (O3,O4) is 0.51 $\mu_\mathrm{B}$. The surface O atoms are coordinated with two nearest Sn atoms, while the sub-surface O atoms are coordinated with three nearest Sn atoms. This difference of crystal geometry leads to larger local induced moments at the  surface O atoms as compared to the moments induced at sub-surface O atoms. The local moments of Li and Sn have negative values, which show that there is an antiferromagnetic type of coupling either between the surface Li-O atoms or sub-surface Sn-O atoms.
The local magnetic moments of Li, Sn, and O atoms are different when Li diffuses to sub-surface and replaces Sn(2) sub-surface atom, see Table~\ref{LMM1}.The magnetic moment induced on each surface O atom (O1,O2) is $\sim0.93 \mu_\mathrm{B}$ per O atom, whereas the moment induced on the each sub-surface O atom (O3,O4) is $\sim 0.68\mu_\mathrm{B}$. The O-2$p$ states are the main source of surface magnetism. Again, Li and Sn atoms have negative induced magnetic moments, which couple antiferromagnetically with the O atoms. Note that LDA+$U$ always gives larger local magnetic moments. When Li is doped at sub-surface Sn(2) site, the local magnetic moments at O, Li, and Sn sites are smaller as compared to the case when Li is doped at surface Sn(1) site. This behaviour is similar to C-doped SnO$_{2}$.~\cite{apl}

\begin{table}[b]
\caption{The calculated local magnetic moments (LMM)(in units of $\mu_\mathrm{B}$ ) of surface and sub-surface atoms when Li is doped at the surface Sn(1) site (right panel) and sub-surface site Sn(2)(left panel) . Values in parentheses show LMM calculated with LDA+$U$}
\begin{tabular}{cc|ccc}
\toprule
Surface atoms &LMM&Surface atoms&LMM \\
\colrule
Li&-0.05(-0.03)&Sn(1)&-0.01(-0.16) \\
O1&1.00(1.05)&O1&0.64(0.93)\\
O2&1.00(1.05)&O2&0.64(0.93)\\
Sub-Surface atoms&&Sub-Surface atom\\\cline{1-4}
  Sn(2)&-0.07(0.07)&Li&-0.07(-0.07)\\
  O3&0.52(0.51)&O3&0.45(0.68)\\
  O4&0.52(0.51)&O4&0.45(0.68)\\
  \botrule
\end{tabular}
\label{LMM1}
\end{table}

\begin{figure}[b]
\centering
\subfigure[]{
\includegraphics[scale=0.4]{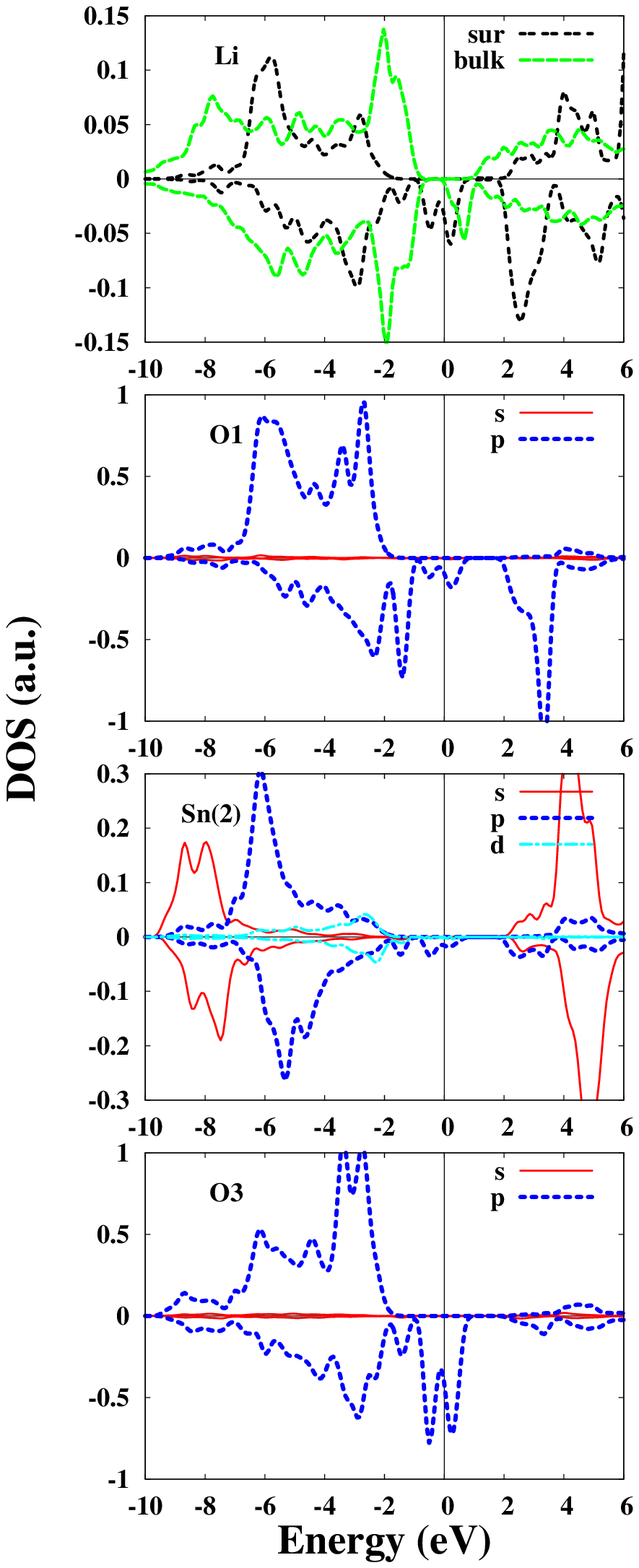}
\label{dos1}}
\subfigure[]{
\includegraphics[scale=0.4]{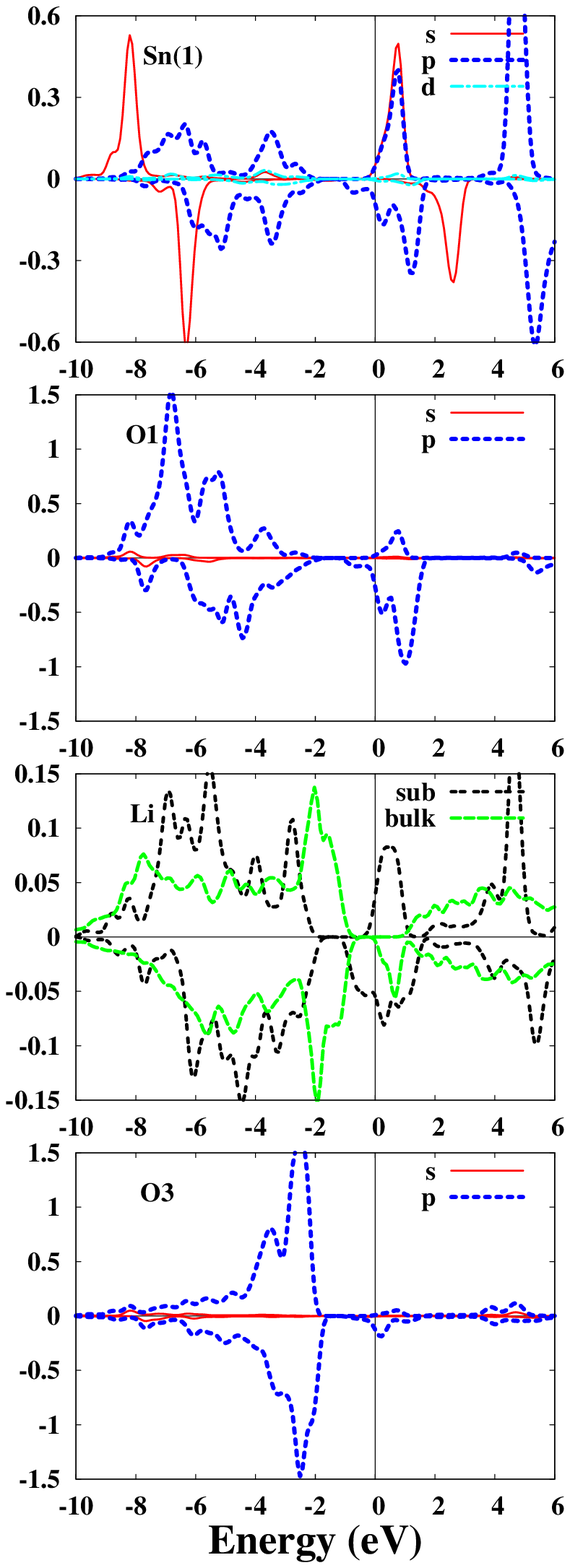}
\label{dos2}}
\label{fig:sursur113dos}
\caption[Optional caption for list of figures]{The LDA+U calculated  projected density of states (PDOS) of Sn, Li, O atoms when Li is doped at (a) surface Sn(1) and (b) sub-surface Sn(2) sites of SnO$_2$ (001).  The positive  (negative) PDOS shows majority (minority) spin states, and the vertical lines show the Fermi level E$_\mathrm{F}$, which is set to zero. Solid (red), dashed (blue), and dotted-dashed(cyan) lines represent $s$, $p$, and $d$ orbitals, respectively. The long-dashed(green) lines represent the bulk Li PDOS and dashed (black) lines show the surface Li PDOS when doped at Sn(1) or Sn(2) site.}
\end{figure}

We have shown that the surface O atoms have larger local magnetic moments than the sub-surface O atoms, and to know the atomic origin of these local moments, we calculated the atom projected density of states (PDOS). Fig.~\ref{dos1} shows the PDOS on the orbitals of the surface and sub-surface atoms when the Li atom is doped at the surface Sn(1) site. Clearly, the Li atom induces magnetism at the (001) surface of SnO$_2$. The low lying $s$ orbitals of Li are spin-polarized and strongly hybridized with the surface $sp$ orbitals of Sn. The Fermi energy (E$_\mathrm{F}$), which is set to zero, is mainly dominated by the $p$ orbitals of O, which indicate that magnetism is mainly induced by the $p$ orbitals and localized at the O atom. 
Majority $s$ spin state of the Li atom is completely occupied and minority spin state is partially occupied leading to a significant spin splitting. The minority surface spin states are driven by strong hybridization of Li with the O1 and O3-$p$ orbitals, which lead to conducting band. These hybridized minority spin states have a large weight at the Fermi energy, whereas the majority spin states have no states at the Fermi level and the majority spins behave as in an insulator. Such 100\% spin-polarized band structure, half-metallic band, is essential for spin based devices. For comparison purpose, we have also shown the PDOS of Li doped at bulk Sn site\cite{arxi}. It is clear to see that surface Li doped system has larger majority spin band gap as compared to the bulk case. The minority surface spin states are formed in the bulk band gap. 
The surface electronic structure is different when Li is doped at sub-surface Sn(2) site [see Fig.~\ref{dos2} ]. The PDOS shows hybridization between the $p$ orbitals of O, and $p$ orbitals of  Sn atom, particularly near E$_\mathrm{F}$. In the majority/minority surface spin band, the surface states are mainly driven by Li-Sn hybridization, and both the bands have no gap at the Fermi energy which shows a metallic behavior. Some of the $p$ states of Sn atoms are also unoccupied, which were occupied when Li was doped at surface Sn(1) site. This partial occupation of $p$ orbitals of Sn also participates in the surface states. A significant spin-polarization of the $s$ electrons of Sn(2) in the valance band is also visible. Such spin polarization of the surface $s$ electrons of Sn (1) is mainly caused by the exchange filed of O1 atoms. The exchange fields of O1 and O3  are smaller than the case when Li was doped at surface Sn(1) site. The PDOS of surface Li has bulk like electronic structure below $-2.0$ eV, however, near the Fermi energy the sub-surface states are formed in the bulk band gap and the Li-Sn hybridized majority spin states destroy the half metallic nature of Li-doped SnO$_{2}$ (001). 

The electronic structures summarize that Li at either surface site induces magnetism, and the magnetism is not strongly localized around the Li atom, but de-localized over Sn and O atoms. This behavior is quite different from C-doped SnO$_2$,~\cite{apl} where magnetism was mainly contributed by the C atom. This different behavior of Li and C in the same host (SnO$_2$) is mainly due to the absence of $p$ orbitals in Li atom. This absence of $p$ orbitals also helps to promote magnetism in bulk Li-doped SnO$_2$, which is again quite different from bulk C-doped SnO$_2$,  where C shows no magnetism in bulk SnO$_2$.\cite{apl,exp}
Note that magnetism either induced by  doped element or cation vacancy (bulk or suface) is mainly contributed by the O atoms surrounding doped element or  cation vacancy.~\cite{arxi,11}  However, the major changes come in the thermodynamic stability of the system, when magnetism is produced by Sn vacancy or  Li.\cite{arxi,11,jmmm} Surface Sn vacancy or Li doped at surface Sn site has lower formation energy than bulk Sn vacancy or Li doped at bulk Sn site.\cite{jmmm, ZnSnO2} 

Finally, to look for the possibility of ferromagnetism in Li-doped SnO2(001), we considered ferromagnetic (FM) and antiferromagnetic (AFM) interactions between the two surface Li atoms by considering a big supercell ($2\times 1\times 3$). We observed that the two Li atoms at surface Sn(1) sites couple ferromagnetically, and FM state is more stable than the AFM state by $\sim 24 \,$meV. 
Such FM coupling mainly occurs between the O atoms due to its large magnetic moment as compared with the local magnetic moments of Li and Sn.  The FM coupling between the O atoms is mediated by the negative spin polarization of the Sn atoms.~\cite{11} Therefore, ferromagnetism is expected in Li-doped SnO$_2$(001) or in  Li-doped SnO$_2$ nonoparticles. 
The experimental reports show that the magnetism of Li-doped SnO$_2$ nanoparticles is neither  governed by defects nor by surface effect because the observed magnetization was not inversely proportional to the nanoparticles size\cite{SnO2-Nano}. We also believe that ferromagnetism in Li-doped SnO$_2$(001) is not induced by surface, but by doping of Li at surface Sn site, consistent with the speculation of Srivastava \textit{et al.}\cite{SnO2-Nano}
Note that Sn vacany induces  magnetism in SnO$_{2}$ and the induced magnetism follow RKKY type interaction.\cite{11}   Usually, light element doped oxides show oscillatory behavior when the interaction between the doped elements is considered at different positions in the supercell.~\cite{marcel,rcs,mgtio2} It is expected that Li-doped SnO$_{2}$ may also follow RKKY interaction.

In summary, 
we investigated the surface magnetism and electronic structures of Li-doped SnO$_2$ (001). The LDA+$U$ calculated formation energy suggested that Li can easily be doped at surface Sn site as compared with bulk and sub-surface Sn sites. The surface relaxation showed 
the surface oxygen atoms were relaxed in the upward direction due to surface strain induced by Li at Sn site, and the sub-surface oxygen atoms were relaxed inward. 
It is shown that Li also induces a large magnetic moment at the SnO$_2$ (001) surface. The magnetic moment, which is localized at the surface and sub-surface atoms, was mainly contributed by O atoms at surface, sub-surface and partially by Sn and Li atoms. Electronic structure calculation showed that Li doped at surface has half metallic character. In the light of our calculations, we predicted that Li-doped SnO$_2$ (001) may be a good material for spin-based devices. We also speculate that Li-doped SnO$_2$ (001) is better than C-doped SnO$_2$ (001) not only due to its low formation energy and magnetism, but also due to half-metallic surface state, which was absent in the C-doped SnO$_2$ (001). Further 
experimental work is required to compare Li-doped  and C-doped SnO$_2$ (001) systems for potential applications in the area of spintronics. 

We are grateful {V\'{\i}ctor M. Garc\'{\i}a-Su\'arez} and S. K. Hasanain for useful discussions.
GR acknowledges the cluster facilities of NCP, Pakistan.

\clearpage

\end{document}